\documentclass[preprint2,longabstract]{aastex}

\newcommand       \mum          {\,{\rm \mu m}}

\shorttitle{}
\shortauthors{Slater et al.}

\begin{document}

\title{Dust Emission from Evolved and Unevolved HII Regions in the
  Large Magellanic Cloud}

\author{C. T. Slater,\altaffilmark{1} M. S. Oey,\altaffilmark{1}
A. Li,\altaffilmark{2}
J.-Ph. Bernard,\altaffilmark{3,4}
E. Churchwell,\altaffilmark{5}
K. D. Gordon,\altaffilmark{6}
R. Indebetouw,\altaffilmark{7,8}
B. Lawton,\altaffilmark{6}
M. Meixner,\altaffilmark{6}
D. Paradis,\altaffilmark{3,4}
W.T. Reach\altaffilmark{9}} 

\altaffiltext{1}{Astronomy Department, University of Michigan,\\ 
  Ann Arbor, MI, 48109}
\altaffiltext{2}{Department of Physics and Astronomy, University of\\
 Missouri, Columbia, MO 65211}
\altaffiltext{3}{Université de Toulouse; UPS-OMP; IRAP;  Toulouse,
  France}
\altaffiltext{4}{CNRS; IRAP; 9 Av. colonel Roche, BP 44346, \\
  F-31028 Toulouse cedex 4, France}
\altaffiltext{5}{University of Wisconsin--Madison, Department of \\
  Astronomy, 475 North Charter Street, Madison, WI 53706}
\altaffiltext{6}{Space Telescope Science Institute, 3700 San Martin
  Drive,\\
 Baltimore, MD 21218}
\altaffiltext{7}{Department of Astronomy, University of Virginia,\\
  P.O. Box 3818, Charlottesville, VA 22903}
\altaffiltext{8}{National Radio Astronomy Observatory, 520 Edgemont\\
  Road, Charlottesville, VA 22903}
\altaffiltext{9}{Universities Space Research Association, Stratospheric \\
Observatory for Infrared Astronomy, NASA Ames\\
 Research Center, MS 211-3, Moffett Field, CA, 94035}

\begin{abstract}
  We present a study of the dust properties of 12 classical and
  superbubble HII regions in the Large Magellanic Cloud. We use
  infrared photometry from {\it Spitzer} (8, 24, 70, and 160$\mum$
  bands), obtained as part of the {\it Surveying the Agents of a
    Galaxy's Evolution} (SAGE) program, along with archival
  spectroscopic classifications of the ionizing stars to examine the
  role of stellar sources on dust heating and processing. Our infrared
  observations show surprisingly little correlation between the
  emission properties of the dust and the effective temperatures or
  bolometric magnitudes of stars in the HII regions, suggesting that
  the HII region evolutionary timescale is not on the order of the
  dust processing timescale. We find that the infrared emission of
  superbubbles and classical HII regions shows little differentiation
  between the two classes, despite the significant differences in age
  and morphology. We do detect a correlation of the 24$\mum$ emission
  from hot dust with the ratio of 70 to 160$\mum$ flux. This
  correlation can be modeled as a trend in the temperature of a
  minority hot dust component, while a majority of the dust remains
  significantly cooler.
\end{abstract}


\section{Introduction}

The life-cycle of interstellar dust remains a largely
under-constrained process. Until recently, attempts to characterize
the evolution of dust have been impeded by uncertainties regarding
dust's composition, physical state, emission processes, and
illuminating radiation field. All of these parameters must be known
prior to drawing any observationally-based conclusion on dust
evolution, and before recent results from the {\it Spitzer Space
  Telescope} emerged, these parameters had been largely uncertain as
well.  A consensus is gradually forming around some of these
properties, with the modeling of the ``unidentified infrared bands''
as vibrational emission from polycyclic aromatic hydrocarbons
\citep[PAHs;][]{leger84,allamandola85}, and models of the size
distribution and cross sections of the PAHs and small grains are now
better defined \citep{li01}.

Many {\it Spitzer} programs have taken advantage of these new dust
models to provide insight on many different classes of objects.
Previous works have characterized the dust content of entire galaxies
\citep[e.g.][]{dale05,draine07} and individual star forming regions
\citep[e.g.][among many others]{compiegne08,berne07}, but consistent
analyses of larger samples of HII regions have been somewhat lacking.
One exception is the work of \citet{gordon08}, which used a
sample of HII regions in M101 to probe metallicity-dependent dust
processing. Their findings linked the ionization parameter of HII
regions with a decrease in the PAH emission feature equivalent widths,
which is evidence for the destruction of PAHs. 

Our work seeks to relate the properties of the stars in HII regions to
their dust content and emission. The radiation output from these
massive, luminous stars dominates the heating of the dust, and 
the stellar SED and luminosity may affect the dust emission.  
Evolved massive stars are also thought to be an important source for
dust production, and stellar wind shocks and supernovae can also be
important for processing dust.  Our work leverages a previously compiled
sample of HII regions with known spectral types for the ionizing
stars. The sample includes regions with ages spanning from 1--5
Myr, including both classical HII regions and evolved, multi-SN
superbubbles.  This range of objects enables us to estimate the dust
processing timescales.  We seek to determine which of these varied
parameters, if any, dominate the dust properties and emission.
In Section \ref{observations}, we describe the observations
and data reduction; Section \ref{analysis} contains our interpretation
and modeling of the observed infrared ratios; Section \ref{stellar}
discusses the relation between stellar properties and infrared
emission; and we explain the applicability of our work to other
observations in Section \ref{discussion}.

\section{SAGE Observations}
\label{observations}

Our sample of H\textsc{II} regions was drawn from the objects in
\citet{ok97}. The sample is listed in Table \ref{stellartable}, with
identifiers from \citet[][; hereafter DEM]{DEM76}.  The sample
contains 8 superbubbles and 4 classical HII regions, and all of the
objects in this sample have spectroscopically-determined spectral
types for their ionizing stars. In Table \ref{stellartable} we have
listed the spectral type of the hottest star in the association, which
dominates the stellar luminosity and spectral energy distribution in
the region.  We have also included the total stellar bolometric
magnitude for the ionizing stars, determined by summing the
luminosities of all stars spectroscopically observed by the work
referenced in the last column. The bolometric corrections had been
applied by the authors of each cited work except for DEM 199, for
which we used the bolometric corrections of \citet{martins05} for O
and B stars and \citet{crowther07} for Wolf-Rayet stars. With the
inclusion of superbubbles, our sample spans a wider range of object
ages, which could potentially demonstrate dust evolution. Many of the
superbubbles also have supernovae and shocks which could drive dust
processing.

The infrared data were obtained from the {\it Spitzer}
Legacy program, Surveying the Agents of a Galaxy's Evolution 
\citep[SAGE;][]{meixner06}. The program observed an area of $\sim7^\circ
\,\times\,7^\circ$, covering the LMC in all IRAC (3.6, 4.5, 5.8, and
8$\mu$m) and MIPS (24, 70, and 160$\mum$) bandpasses. The
individual observations were calibrated, combined, and mosaicked by the
SAGE team, with a pipeline similar to that of the GLIMPSE survey
\citep[see][for further details]{meixner06,benjamin03}.

\subsection{Photometry}
                                   
In order to compare surface brightnesses of the target objects at
different wavelengths it is necessary to convolve the images to the
lowest resolution of the observations, which in our case is the 160
$\mu$m band (full-width at half-maximum of $40^{\prime\prime}$). Due
to diffractive features in the instrument's point-spread function
(PSF), such as Airy rings and spikes, a convolution with a Gaussian
kernel would not produce sufficiently accurate results. Instead we
used the convolution kernels described in \citet{gordon08}, which are
calculated from model PSFs. This convolution also alleviates the need
for aperture corrections when computing band ratios, since the
aperture correction will be a constant factor for all bands.

We followed the prescription of \citet{helou04} for using the
3.6$\mum$ images to remove the stellar flux from the 8 and 24$\mum$
images. We computed the non-stellar flux ($F_\nu^{\rm ns}$) from the
observed flux ($F_\nu$) in each band using
\begin{equation}
F_\nu^{\rm ns}(7.9\mum) = F_\nu(7.9\mum) - 0.232 F_\nu(3.6\mum)
\end{equation}
and
\begin{equation}
F_\nu^{\rm ns}(24\mum) = F_\nu(24\mum) - 0.032 F_\nu(3.6\mum) ~, 
\end{equation}
where the coefficients are from \citet{helou04}. These constants were
determined by comparing the mid-infrared emission of stellar
population synthesis models to their 3.6$\mum$ emission. The infrared
emission of these models is not strongly sensitive to star formation
history or metallicity and should be generally applicable, but they
are still approximate. Using the 3.6$\mum$ images for this subtraction
also assumes that the 3.6$\mum$ band includes no dust emission, even
though a small amount of dust emission is often present in this band.
A visual check of the subtraction confirmed the efficacy of the
subtraction at removing numerous faint point-sources from the
images. Since the contribution from stellar sources at 71 and 160
$\mu$m is negligible, we performed no correction on those bands.  When
discussing the 8$\mum$ and 24$\mum$ photometry in what follows, we
will be referring to these corrected non-stellar fluxes.

We performed aperture photometry on the objects, with apertures
defined by hand. In all cases, we set the apertures such that they
included all of the H$\alpha$ flux that could reasonably be associated
with the target object. The Magellanic Cloud Emission-line Survey
\citep[MCELS;][]{smith98} H$\alpha$ imaging of the LMC was used for
this comparison.  Figure \ref{SBimages} shows infrared and H$\alpha$
images of all of the superbubbles in our sample, and Figure
\ref{NonSBimage} shows the classical HII regions. In both figures the
photometric apertures are outlined in white.

\subsection{Background and Error Estimation}

Our photometry is potentially contaminated by both diffuse emission
from the Milky Way and by emission from unrelated dust in the LMC in
the foreground or background of our objects (hereafter collectively
referred to as ``background'' emission). We subtracted this background
emission from our objects, measuring the background level with an
annulus spanning 14 to 17 arcminutes from the center of each
region. We used the mode (the most commonly occurring value after
binning the values into 0.01 MJy/steradian bins) of the pixels in this
annulus as the background level. The mode was chosen rather than the
mean in order to reduce the sensitivity of the background estimation
to bright structures from neighboring clouds. We note that it is
impossible to completely account for emission unrelated to our objects
that may appear within our target apertures, since much of the dust
emission varies on size scales smaller than our apertures. Though this
should not cause a systematic error in our photometry, it will still
necessitate a level of caution in interpreting values for specific
target objects.

Given that it is possible for this background contamination to exist
despite our efforts to remove it, we would like to know the extent to
which it may compromise our measurements. We therefore need to know
the variance we would measure with our apertures {\it if they were
  placed randomly} and not on our target objects. By placing a number
of ``error estimation'' apertures of the same size as our object
apertures and measuring the variance in integrated light amongst them,
we can thus obtain some estimate of the uncertainty in our
photometry. This still does not allow us to measure the background's
contribution to any single object's photometric measurement any better
than we did with the background annulus, but it does inform us of the
uncertainty in that measurement.  The flux densities for all of our
objects, corrected for stellar and background emission, are listed in
Table \ref{phottable}. These flux densities have not had an aperture
correction applied, so they are likely to be systematically low. This
does not affect the band ratios used in this work though, since the
necessary correction factors are identical for all bands as a result
of convolving each image to a common PSF. Contamination by unremoved
background is the dominant source of uncertainty in our photometry and
the error bars on the figures and in Table \ref{phottable} reflect
that.

\section{Dust Emission and Modeling}
\label{analysis}

To allow for comparisons between objects with different 
luminosities, we constructed three different flux ratios from the data,
following the methods of \citet{DL07}. These are
\begin{eqnarray}
P_{7.9} & = & \frac{\nu F_\nu^{\rm ns}(7.9\mum)}{\nu F_\nu(71\mum) + 
              \nu F_\nu(160\mum)} , \\
P_{24} & = & \frac{\nu F_\nu^{\rm ns}(24\mum)}{\nu F_\nu(71\mum) + 
              \nu F_\nu(160\mum)} , 
\end{eqnarray}
and
\begin{equation}
R_{71} = \frac{\nu F_\nu(71\mum)}{\nu F_\nu(160\mum)} .
\end{equation}
$P_{7.9}$ is primarily a tracer of the aromatic emission commonly
associated with PAHs, while $P_{24}$ tends to trace hot, thermalized
dust. Both of these are normalized by the total thermal dust emission,
traced by the quantity $\nu F_\nu(71\mum) + \nu F_\nu(160\mum)$. 
The ratio $R_{71}$ measures the temperature of the dust that has
reached thermal equilibrium, which is generally comprised of larger
grains with radii greater than 100 \AA. Since the temperature at which
these grains equilibrate is determined solely by the amount of
radiation they receive, the 70/160$\mum$ ratio is also useful as a
tracer of the total radiation content of the region. 

Figure \ref{corr8and24} shows the effect of radiation intensity on
both the PAH and VSG populations. The top panel shows that $P_{7.9}$ is
largely uncorrelated with the total radiation content of the region.
This is expected as the PAH emission is the result of stochastic
heating of grains: the level of the PAH emission is directly
proportional to the radiation intensity, but the overall PAH emission
spectral shape does not vary with the radiation intensity
\citep{DL01}.  Since the ratio $P_{7.9}$ is normalized by the
far-infrared emission ($\nu F_\nu(71\mum) + \nu F_\nu(160\mum)$) which
is also proportional to the radiation content of the region, the ratio
is therefore largely independent of the intensity of the radiation.

The lower panel in Figure \ref{corr8and24} indicates that $P_{24}$ ratio
increases with radiation intensity. Since the 8$\mum$ data show that
PAH emission is largely uncorrelated with the 70/160$\mum$ ratio,
it is clear that the trend in 24$\mum$ must be the result of
thermalized dust, rather than the PAH emission within the 24$\mum$
bandpass. At low temperatures, the thermal dust emission is detected
only in the 70 and 160$\mum$ bands, but for dust exposed to high
radiation intensities, the thermal peak can reach into the 24$\mum$
band. We therefore hypothesize that as the overall luminosity of
the region increases, the peak radiation intensity seen by the dust
and consequently the temperature of the dust also increase in a
correlated way.

Which parameter determines the position of an HII region along this 24
$\mu$m sequence, and which parameters determine the locus of the
sequence itself?  Since we hypothesize that this trend is driven by
the effects of the illuminating radiation field, and not a change in
dust composition, we examine parameterizations of the radiation that
can best reproduce this correlation.

Our simple models of the sequence in 24$\mum$ emission are shown in
Figure \ref{modelfit24}. In this figure we have plotted 
$P_{24} - 0.14 P_{7.9}$, which was suggested by \citet{draine07} 
as a way to remove the contribution of PAHs to the 24$\mum$ band, 
and leave only the flux from thermalized grains.  
This minimizes any effects caused by variations in PAH abundances.  
Figure~\ref{modelfit24} shows the same tight correlation in this 
measure of the dust emission as that for $P_{24}$ alone, 
uncorrected for PAH emission.

Using absorption cross-sections and grain size distributions from
\citet{DL07}, we constructed a variety of models to compare to our
observed sequence. Our models compute only the emission from the
thermalized dust. In thermal equilibrium the dust grain emission must
equal absorption, hence
\begin{equation}
\int_0^{\infty} C_{abs}(a, \lambda) u_\lambda \, d\lambda = \frac{4 \pi}{c}
\int_0^\infty C_{abs}(a, \lambda) B_\lambda(T) \,d\lambda, 
\label{equilibriumEqn}
\end{equation}
where $C_{abs}(a,\lambda)$ is the absorption cross-section as a
function of grain radius $a$ and wavelength $\lambda$, $u_\lambda$ is
the radiation energy density, and $B_\lambda(T)$ is the Planck
function \citep{li01}.  We follow \citet{DL07} in parameterizing the
radiation field as $u_\lambda = U u^{\rm MMP83}_\lambda$, where we use $U$
as a scaling coefficient on the mean interstellar radiation field
$u_\lambda^{\rm MMP83}$ from \citet{MMP83}. This parameterization was
chosen primarily to aid comparison to other works; since the grain is
assumed to reach an equilibrium temperature, only the energy content
of the radiation field affects the dust emission and not the spectral
energy distribution. For a given radiation field, we compute the
temperature $T$ for each grain size such that the equation of
equilibrium (Equation \ref{equilibriumEqn}) is satisfied, then use
that temperature to compute the infrared emission of the dust
grain. We then sum this emission with the emission from grains of
other sizes, using the size distribution from \citet{li01}.

In the simplest model, all of the dust in the HII region is heated by
a single radiation intensity. We then computed a set of models
spanning a range of radiation intensities ($U$), shown by the red,
dotted line in Figure \ref{modelfit24}. This simple model shows that
as the parameter $U$ increases, the model's position in the
$R_{71}-P_{24}$ plot moves towards the upper right.  This trend
follows a similar slope to that of our observed set of HII regions,
but the models exhibit a 70/160$\mum$ ratio that is much higher than
those of the observed regions. This suggests that though some dust
must be exposed to high radiation intensities to reproduce the
24$\mum$ emission, cold dust is also needed to reproduce the observed
low 70/160$\mum$ ratios. These dust components could be physically
separate and have distinct temperatures, or they could form a
continuum spanning a range of temperatures, but dust must be present
at both high and low temperatures.

We therefore extend our models to include a mass fraction $\gamma$ of
dust exposed to high radiation intensities, and a mass fraction $(1 -
\gamma)$ that is exposed to a fixed low radiation intensity $U_{\rm
  min}$. We have adopted this parameterization from \citet{DL07}, but
note that we do not include the power law distribution of radiation
intensities that was used in their work; each of the dust components
sees only a single radiation intensity.  Namely, the dust is heated by
two radiation intensities, with $\left(1-\gamma\right) U_{\rm min}$ of
the total emission arising from the cold dust (heated by $U_{\rm
  min}$) and $\gamma U$ of the total emission arising from the hot
dust (heated by $U$).  We ran a grid of models over this parameter
space ($\gamma$, $U_{\rm min}$) for a range of $U$ and sought to
minimize the chi-square between the models and the observations.  The
best-fitting model had the parameters $\gamma = 0.0015$ and $U_{\rm
  min} = 0.4$, shown by the blue, dashed line in Figure
\ref{modelfit24} and with tick marks indicating the position of
$U=200$ and $U=800$ along the sequence. The small value of $\gamma$
required to fit the observations suggests that the bulk of the dust
inside our apertures is relatively cold, potentially originating from
the photodissociation region surrounding the ionized gas. This cold
dust is at temperatures of roughly $15$ K, but grains of different
sizes equilibrate at different temperatures and thus there will be a
range of dust temperatures. For comparison, grains exposed to $U=600$
equilibrate between roughly $40$ and $55$ K.

It is important to remember that these models are merely suggestive,
and cannot be considered a definitive explanation of the observed
trend in $P_{24}$. We have here sought to explain the observations in
terms of a single parameter model, where the only difference between
objects is the maximum radiation intensity seen by a small fraction of
the dust. It is worth noting that although we do not consider the
derived parameters ($\gamma$, $U_{\rm min}$, $U$) to precisely
represent the actual radiation intensities, the $U$ values inferred
from this simple model (i.e. $U>100$) are indeed consistent with the
premise that the 24$\mum$ emission is mainly from dust attaining
equilibrium temperatures. As shown in Figure 13 and Figure 15 of
\citet{DL07}, at $U>100$ the 24$\mum$ emission is dominated by dust
with equilibrium temperatures. The fact that such a simple model is
able to reproduce the observed trend is certainly interesting, but
should not be taken as a conclusive determination of the radiation
field illuminating these objects. Our model is merely useful for
illustrating that the observed sequence is set by the hottest dust
component, and the temperature that this dust reaches determines the
HII region's position along the sequence. Similarly, it is beyond the
power of our data to suggest exactly what amount of dust is in the hot
or cold components of the HII region itself since we could not
definitively remove unrelated background or foreground emission from
our measurements.

\section{Effects of Stellar Properties}
\label{stellar}

Figure \ref{sptype} shows our three infrared ratios plotted
against the spectral type of the hottest ionizing star and the total
bolometric magnitude of all the stars in the region. The top two
panels show again that there is little systematic change in the PAH
emission between objects. The center and lower panels can be used to
diagnose the dust heating. What the data show is that neither the peak
temperature nor the average temperature of the dust changes
significantly as a function of stellar spectral type or total region
luminosity. 

This is not to say that the total dust emission is completely
unaffected by the stellar luminosity; since they are both ratios they
do not probe the total dust luminosity.  Instead the 70/160$\mum$
ratio tells us that in regions with higher stellar luminosities, the
dust does not have systematically higher temperatures. Looking at the
effect of the bolometric luminosity on the 24$\mum$ emission in the
middle panel of Figure \ref{sptype}, we see that the stellar
luminosity also does not influence the peak temperature of the
dust. In Table \ref{stellartable} we have computed $L_{\rm IR}/L_{\rm
  bol}$, which is the ratio of the measured infrared emission
(assumming 24, 70, and 160$\mum$ bandpasses, but without correcting for
emission outside these bands) to the bolometric luminosity of the
stars in the region. Most of the regions have values of $L_{\rm
  IR}/L_{\rm bol}$ between 5 and 15\%, though this is a lower bound
and integrating over the entire infrared spectrum will yield higher
values. The fairly constant value supports our picture that the total
infrared emission necessarily increases to accommodate increased
stellar emission but, as shown above, the dust temperature does not
appear to be strongly affected.

We can estimate the temperature difference that would be expected from
the range of bolometric luminosities spanned by our objects, using
basic radiative equilibrium (Equation \ref{equilibriumEqn}), which
implies that the stellar luminosity and dust temperature are related
as $L_\star \propto T_{\rm dust}^4$, if the dust is like a black
body. But because the far infrared dust absorption cross-section is
approximately $C_{abs} \propto \lambda^{-2}$, the dust temperature is
actually related to the stellar luminosity by $L_\star \propto T_{\rm
  dust}^{6}$. Our objects span roughly a factor of 100 in luminosity,
which corresponds to a factor of 2 increase in dust temperature. We
therefore would only expect a modest trend in $70/160$ $\mum$, which
would be further obscured by the large intrinsic scatter between
objects. Note that this calculation also assumes that the dust in all
objects is located the same distance from the illuminating stars. If
the dust in more luminous regions was systematically closer or further
from the stars, then the relative dilution of the radiation field may
also effect emission.

\section{Discussion}
\label{discussion}

It is important to highlight the fact that the correlation between the
70/160$\mum$ ratio and 24$\mum$ emission we have observed exists
despite the numerous parameters of the HII regions that could affect
their dust content and emission. Objects in our sample cover a wide
range of morphologies and illuminating stellar sources. This could
generate variations in the dust temperature distribution and resulting
infrared emission, but this is not observed. Instead, the objects sit
on a well-defined sequence with only minimal modification by geometric
effects.

By including in our sample both classical HII regions and more evolved
superbubbles, our sample also spans a range of ages. The superbubbles
are generally older HII regions that have had time to evacuate a shell
of material, and it would be reasonable to expect the dust in these
objects to have undergone more processing than their younger
counterparts.  Additionally, superbubbles are often old enough for
supernovae to have occurred within them, which could affect the dust
content. We do not observe any evidence of such effects, though, since
in both 8$\mum$ and 24$\mum$ the two classes of objects exhibit no
clear differentiation from each other. This could be a consequence of
our significant photometric uncertainties, but is compatible with
grain processing being completed on timescales shorter than the
typical age of our classical HII regions (1--2 Myr), or on timescales
longer than the typical age of the superbubbles (4 -- 5 Myr).

Our results can also explain observations of HII regions in other
galaxies. Galaxies in the {\it Spitzer} Infrared Nearby Galaxies
Survey (SINGS) have exhibited a decreased 8/24$\mum$ ratio in clumps
corresponding to large HII regions, while the 8/160$\mum$ ratio
remains fairly smooth \citep{bendo08}. The 8/24$\mum$ ratio is roughly
a factor of two less in the clumps than it is in the rest of the
galaxy, a factor which can be easily accounted for by our 24$\mum$
observations. Our models therefore imply that the observed clumping
in 8/24$\mum$ can be explained by higher dust temperatures in these
regions, without having to invoke variation in dust abundance.
Another study of SINGS galaxies by \citet{calzetti07} found that in
the integrated light of galaxies there is correlation between 24
$\mum$ and Paschen $\alpha$ emission, which suggests that the 24$\mum$
emission is indeed related to star formation. Similarly, starburst
galaxies also show enhanced 24$\mum$ emission that is attributed to
hotter dust (e.g., Hanish et al. 2010).

Since our work deals only with the integrated light from the target
objects, we are not able to distinguish between emission from the
ionized region and emission from the photodissociation region
(PDR). Emission from PAHs has typically been found within the PDR and
not the ionized region \citep{kassis06}, and consequently the
properties of the HII region and the PAH emission may not be closely
linked. We are also unable to say if the cold dust emission we
observed is from the ionized region or the PDR. With its reduced
radiation intensity, the PDR would be the logical source for the cold
dust emission, but emission from the ionized region could be present
as well.

\citet{gordon08} find a decrease in the PAH equivalent widths for HII
regions as a function of their ``ionization index'', which is
dominated by the ratio of [Ne III]/[Ne II] emission lines. The
ionization index is directly linked to the temperature of the ionized
gas, which could be responsible for the dust processing. We should see
a similar effect in comparing our superbubbles, which have low
ionization parameters due to geometric dilution of the radiation
field, with our classical HII regions. But as described above, we find
no systematic differences in 8$\mum$ emission between the two classes
of objects. Our objects span a comparable range of ionization
parameters to the \citet{gordon08} objects, and it is thus surprising
that all of our objects exhibit so little variation in $P_{7.9}$.

\section{Conclusions}
\label{conclusions}

We find that the correlation between the 70/160$\mum$ ratio and 24
$\mum$ emission is seen in both HII regions and superbubbles. We have
qualitatively reproduced this trend with a simplified dust model that
interprets this effect purely as the result of variations in the dust
temperature between objects. That is, increasing an object's dust
temperature alters its position in the $P_{24}-R_{71}$ diagram (Figure
\ref{modelfit24}) along the trend we observe. We therefore cannot
infer grain processing or variation in grain composition.

Additionally, we do not detect any correlation between the dust
emission ratios and the spectral type of the ionizing stars in the
regions or the total luminosity of all stars in the region. There is
also no differentiation in infrared emission between classical and
superbubble regions, despite the significant differences in age,
morphology, and ionization. This suggests that we are not observing
processing or composition differences on the evolutionary timescale
for superbubbles.

\acknowledgments 

This research is based on observations made with the
Spitzer Space Telescope, which is operated by the Jet Propulsion
Laboratory, California Institute of Technology under contract with
NASA.  Support for this work was provided by NASA through an award
issued by JPL, and by NSF grant AST-0806476. We would like to thank
K. Sellgren and A. Witt for helpful discussions, along with the
anonymous referee for their insightful comments.

\clearpage

\onecolumn
\begin{table}
\begin{center}
\caption{Stellar Parameters for Target Objects\label{stellartable}}
\begin{tabular}{l c c c c c c}
\tableline\tableline 
& Superbubble & Hottest ionizing & Total bolometric & $n_{\star}\tablenotemark{a}$ & $L_{IR}/L_{bol}$ & Reference \\
Name & or classical & star & magnitude & & &\\
\tableline 
DEM 25 & S & O9 V & $-8.88$ & $10$ & $0.058$ & \citet{oey96a}\\
DEM 31 & S & WN6 & $-11.6$ & $24$ & $0.011$ & \citet{oey96a}\\
DEM 34 & C & O3 III & $-13.1$ & $49$ & $0.029$ & \citet{parker92}\\
DEM 50 & S & O6.5 V & $-10.7$ & $12$ & $0.073$ & \citet{oey96a}\\
DEM 106 & S & O6.5 V & $-11.4$ & $16$ & $0.055$ & \citet{oey96a}\\
DEM 152 & S & O5 III & $-12.4$ & $64$ & $0.087$ & \citet{oey95}\\
DEM 192 & S & WC5 & $-12.3$ & $53$ & $0.0071$ & \citet{oey98}\\
DEM 199 & C & WR & $-11.8$ & $35$ & $0.15$ & \citet{garmany94}\\
DEM 226 & S & O6.5 V & $-10.5$ & $10$ & $0.091$ & \citet{oey96a}\\
DEM 243 & C & O9 V & $-11.0$ & $17$ & $0.091$ & \citet{oey96a}\\
DEM 301 & S & O3 I & $-10.8$ & $24$ & $0.055$ & \citet{oey96a}\\
DEM 323 & C & O3 III & $-12.2$ & $27$ & $0.13$ & \citet{massey89}\\

\tableline
\end{tabular}
\tablenotetext{a}{The number of stars whose luminosities are
  summed to compute the total bolometric magnitude.}
\end{center}
\end{table}

\begin{table}
\begin{center}
\caption{Measured Infrared Fluxes for Target Objects\label{phottable}}
\begin{tabular}{l c c c c c c c c}
\tableline\tableline 
& IRAC4 & MIPS24 & MIPS70 & MIPS160  \\
& 8.0$\mum$ & 24$\mum$ & 70$\mum$ & 160$\mum$ \\
Name & (Jy) & (Jy) & (Jy) & (Jy) \\
\tableline 
DEM 25 & $2.2$ $\pm$ $4.5$ & $4.3$ $\pm$ $2.5$ & $87$ $\pm$ $69$ & $220$ $\pm$ $200$ \\
DEM 31 & $2.9$ $\pm$ $1.9$ & $3.4$ $\pm$ $1.4$ & $230$ $\pm$ $39$ & $520$ $\pm$ $89$ \\
DEM 34 & $1.4$ $\pm$ $16$ & $1.6$ $\pm$ $13$ & $35$ $\pm$ $360$ & $90$ $\pm$ $1000$ \\
DEM 50 & $28$ $\pm$ $4.8$ & $28$ $\pm$ $2.3$ & $690$ $\pm$ $48$ & $1300$ $\pm$ $130$ \\
DEM 106 & $45$ $\pm$ $2.3$ & $110$ $\pm$ $2.0$ & $870$ $\pm$ $42$ & $1600$ $\pm$ $100$ \\
DEM 152 & $150$ $\pm$ $8.6$ & $530$ $\pm$ $9.9$ & $3900$ $\pm$ $160$ & $4300$ $\pm$ $350$ \\
DEM 192 & $6.0$ $\pm$ $1.4$ & $23$ $\pm$ $1.2$ & $320$ $\pm$ $25$ & $390$ $\pm$ $63$ \\
DEM 199 & $130$ $\pm$ $24$ & $440$ $\pm$ $52$ & $3800$ $\pm$ $660$ & $5300$ $\pm$ $1100$ \\
DEM 226 & $24$ $\pm$ $15$ & $37$ $\pm$ $18$ & $760$ $\pm$ $300$ & $1200$ $\pm$ $600$ \\
DEM 243 & $37$ $\pm$ $6.8$ & $84$ $\pm$ $3.6$ & $1100$ $\pm$ $79$ & $1900$ $\pm$ $230$ \\
DEM 301 & $20$ $\pm$ $3.3$ & $30$ $\pm$ $1.8$ & $480$ $\pm$ $57$ & $1100$ $\pm$ $140$ \\
DEM 323 & $120$ $\pm$ $100$ & $230$ $\pm$ $100$ & $3800$ $\pm$ $1600$ & $9700$ $\pm$ $4300$ \\

\tableline
\end{tabular}
\end{center}
\end{table}

\begin{figure}
\epsscale{0.90}
\plotone{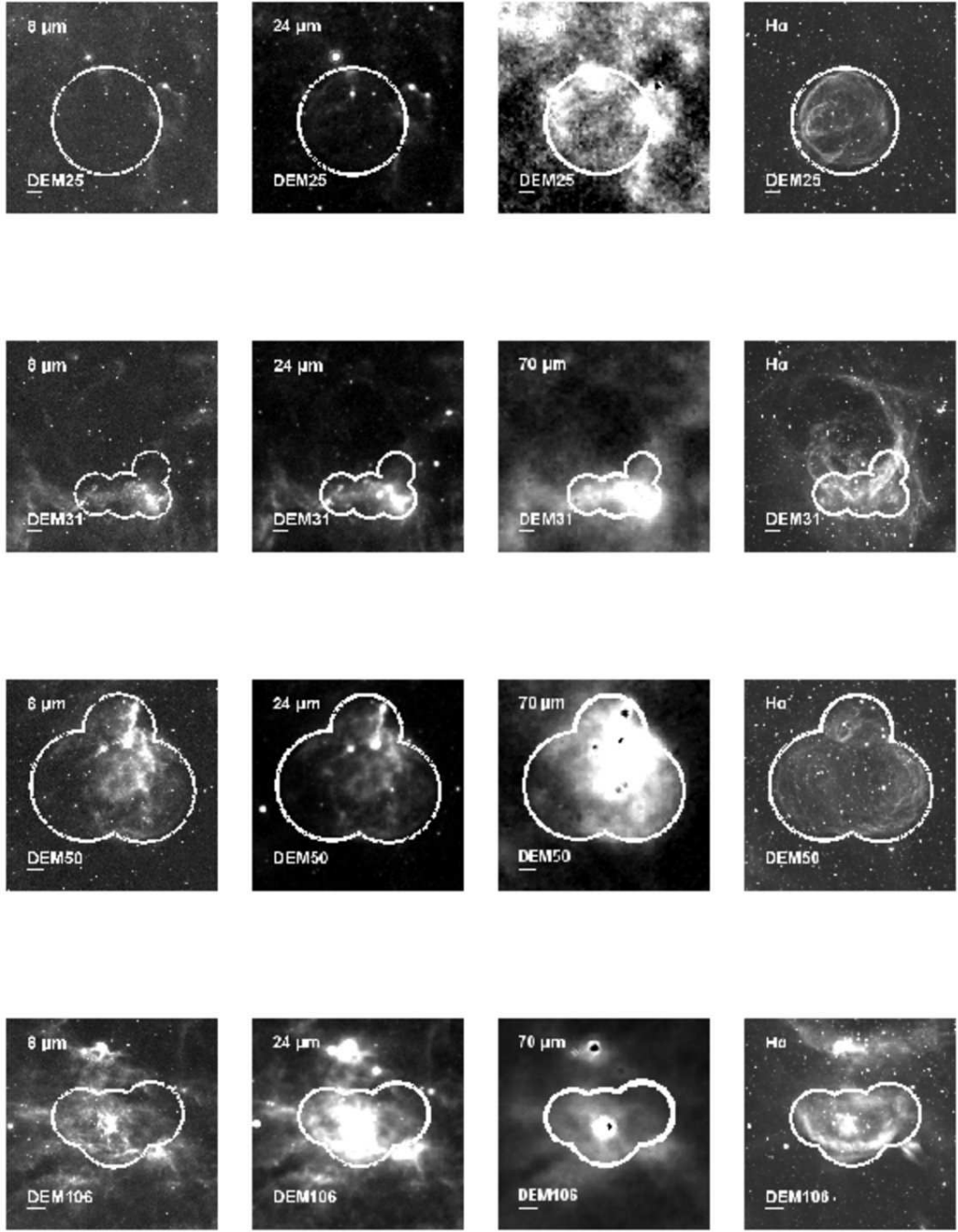}
\caption{Images of the superbubbles in our sample in $8$, $24$,
  $70\mum$ and H$\alpha$, with our object apertures overlaid. The
  scale bar in the lower left of each image is one arcminute,
  corresponding to $15\,{\rm pc}$ at the distance of the LMC
  \citep[$52\,{\rm kpc}$,][]{szewczyk08}.
\label{SBimages}}
\end{figure}

\setcounter{figure}{0}
\begin{figure}
\epsscale{0.90}
\plotone{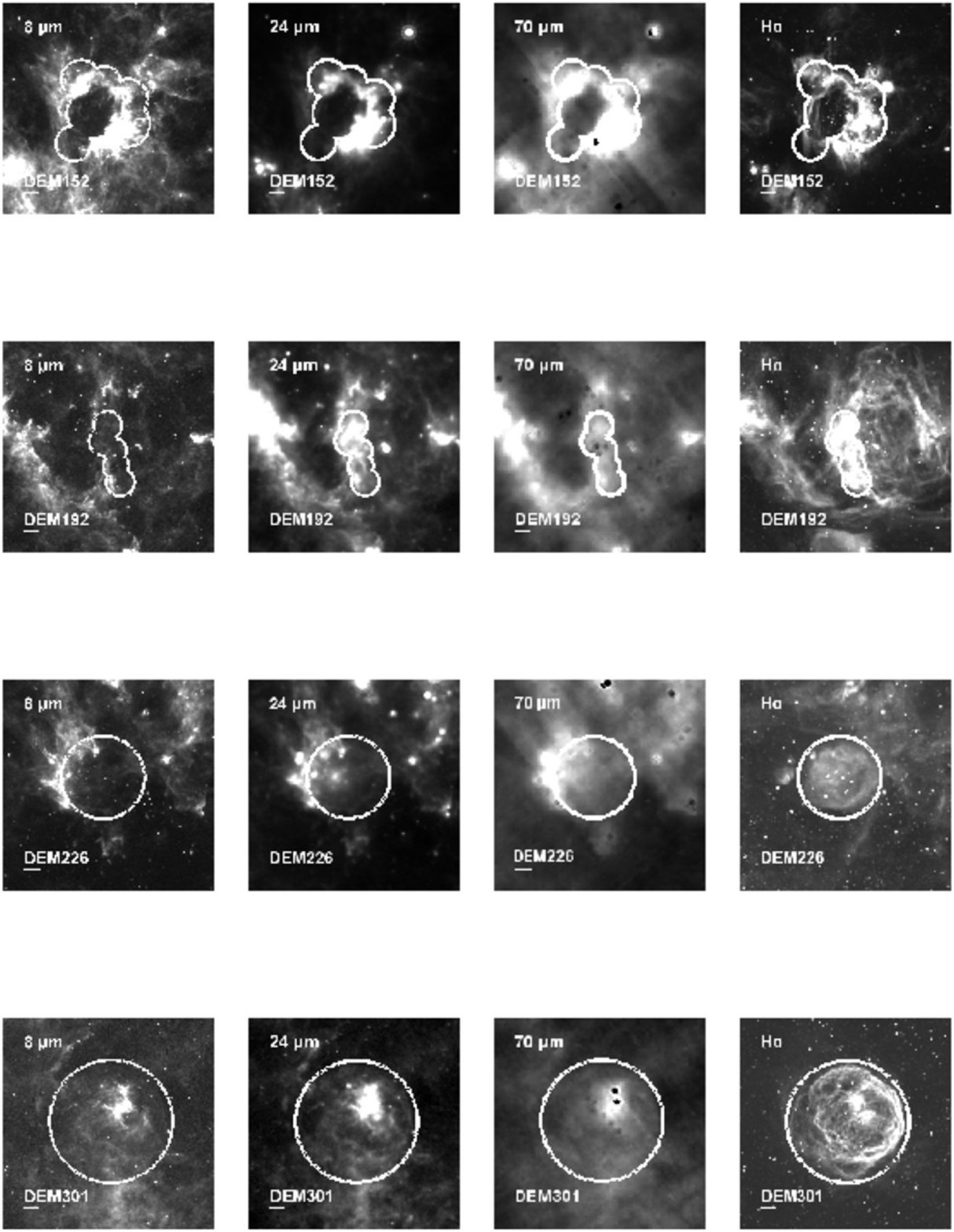}
\caption{continued.}
\end{figure}

\begin{figure}
\epsscale{0.90}
\plotone{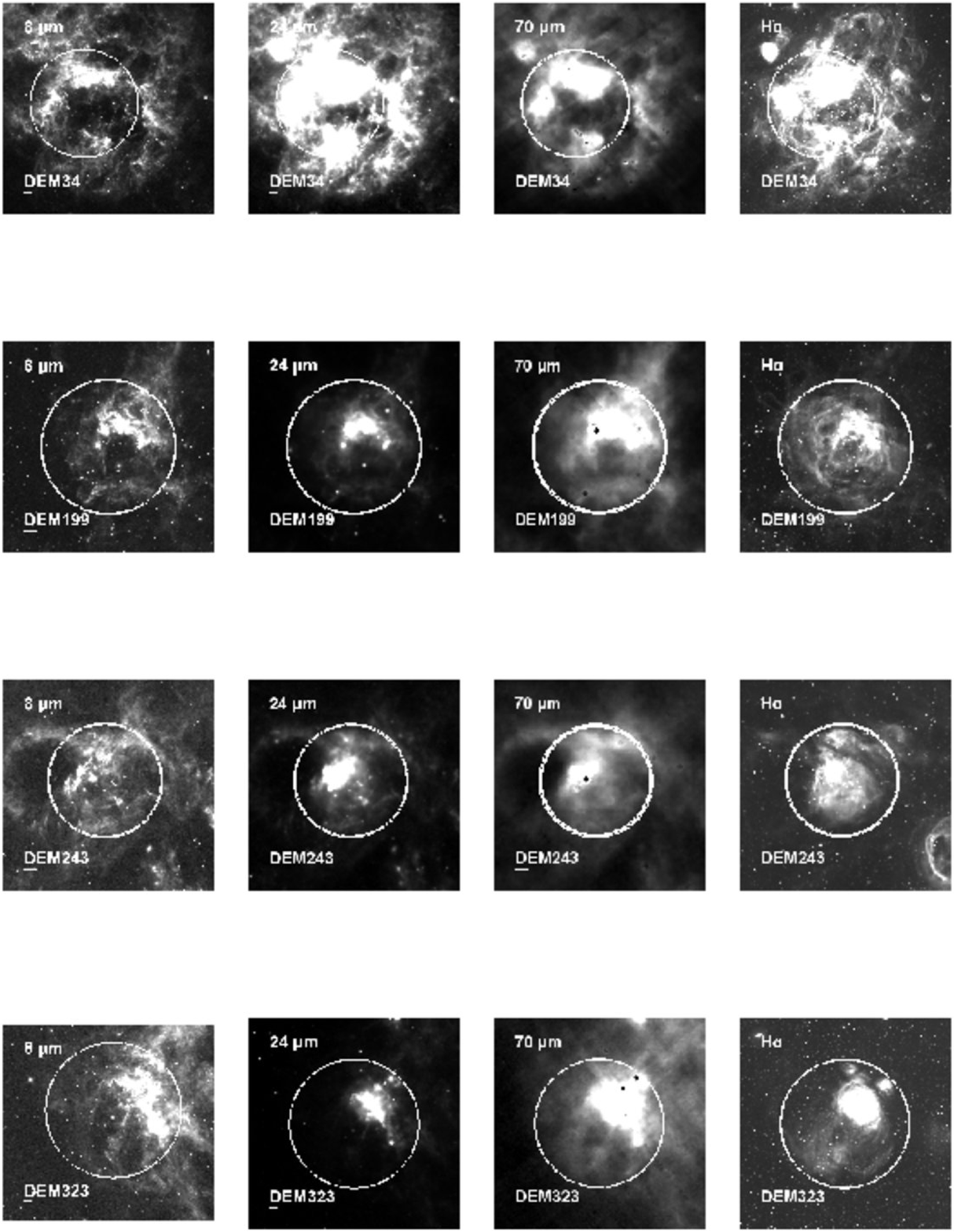}
\caption{Same as Figure \ref{SBimages}, but showing the classical
  HII regions in the sample.\label{NonSBimage}}
\end{figure}

\begin{figure}
\epsscale{0.90}
\plotone{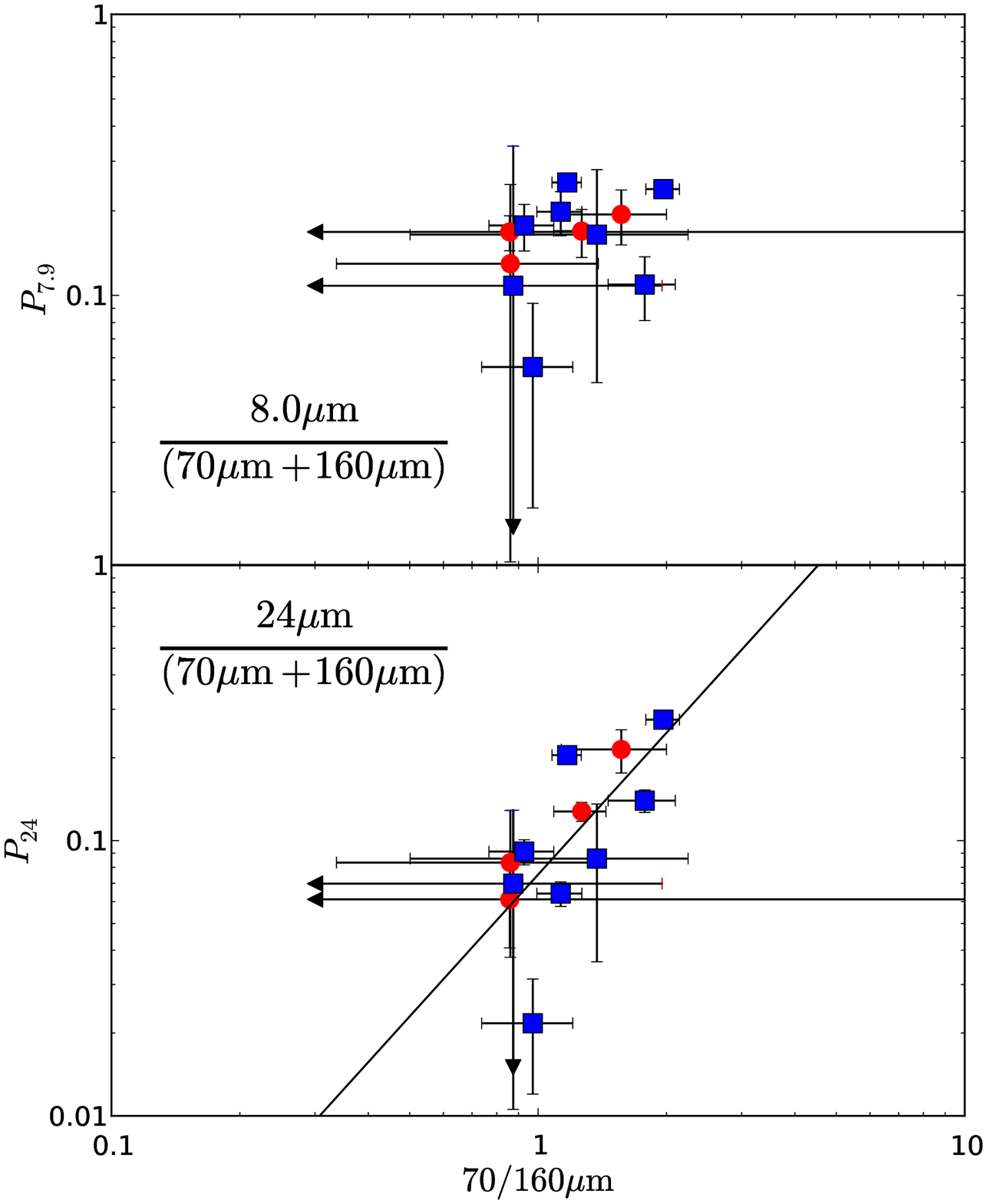}
\caption{{\it Top}: PAH emission as a function of radiation intensity
  of the region, showing little correlation. Blue squares are
  superbubbles, while red circles are classical HII regions. {\it
    Bottom}: 24$\mum$ emission shows a clear increase as the
  radiation content of the region increases.  Symbols are
  as above, and the solid line is a fit to all points. \label{corr8and24}}
\end{figure}

\begin{figure}
\epsscale{1.00}
\plotone{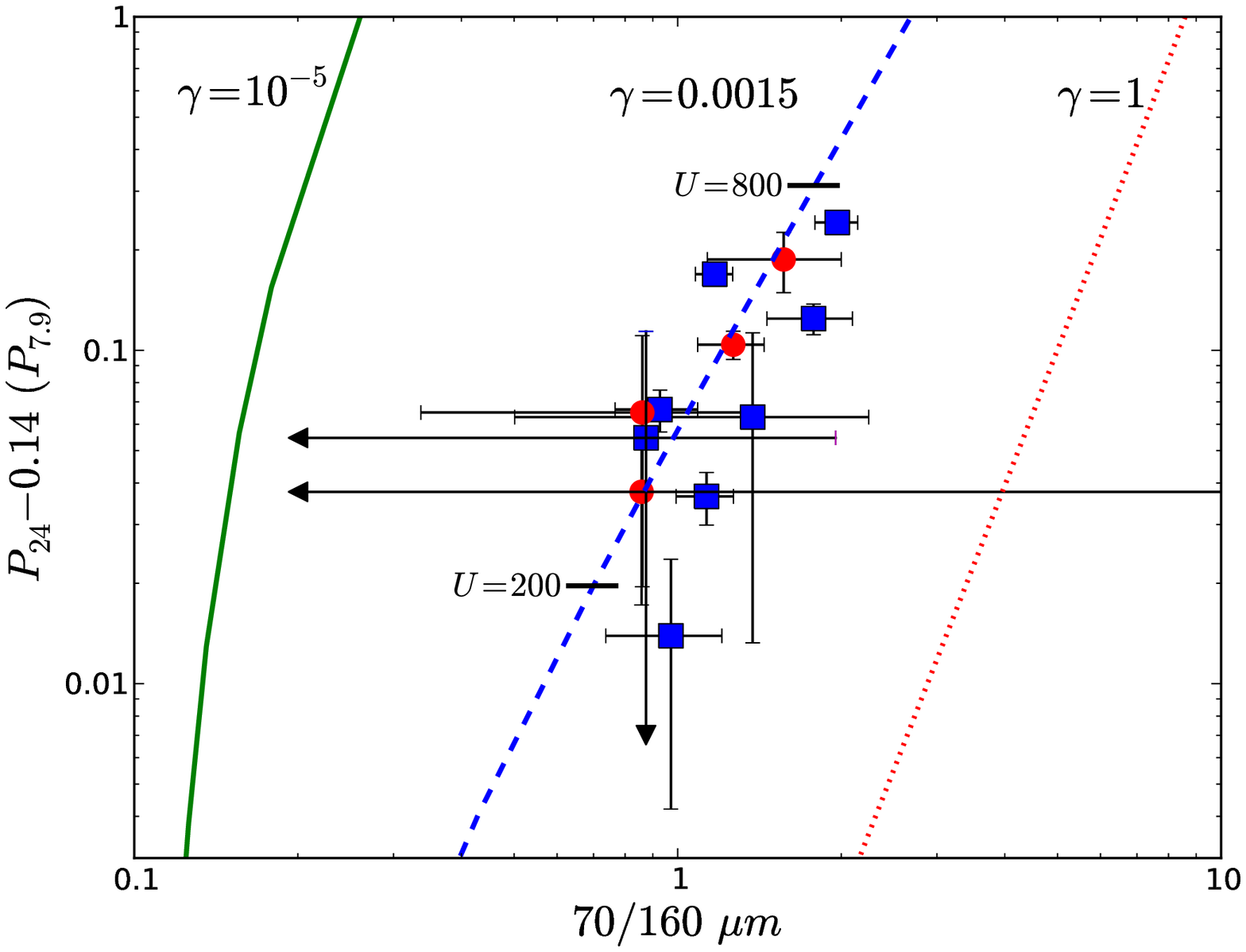}
\caption{The 24$\mum$ thermal emission component shown against the
  70/160$\mum$ ratio. Superbubbles are blue squares, and classical
  HII regions are red circles. The best-fit dust model is shown by the blue
  dashed line, which contains a small fraction of hot dust determined
  by $\gamma$. The red dotted line shows a model where all of the dust
  is hot $\gamma=1$, and the green solid line shows a model with only
  a miniscule fraction of hot dust ($\gamma=10^{-5}$). Tick marks are
  shown on the best-fit model at $U=200$ and $U=800$ to give an idea
  of the range of radiation intensities ($U$) needed to span the
  observations. \label{modelfit24}}
\end{figure}

\clearpage
\thispagestyle{empty}
\begin{figure}
\epsscale{1.00}
\plotone{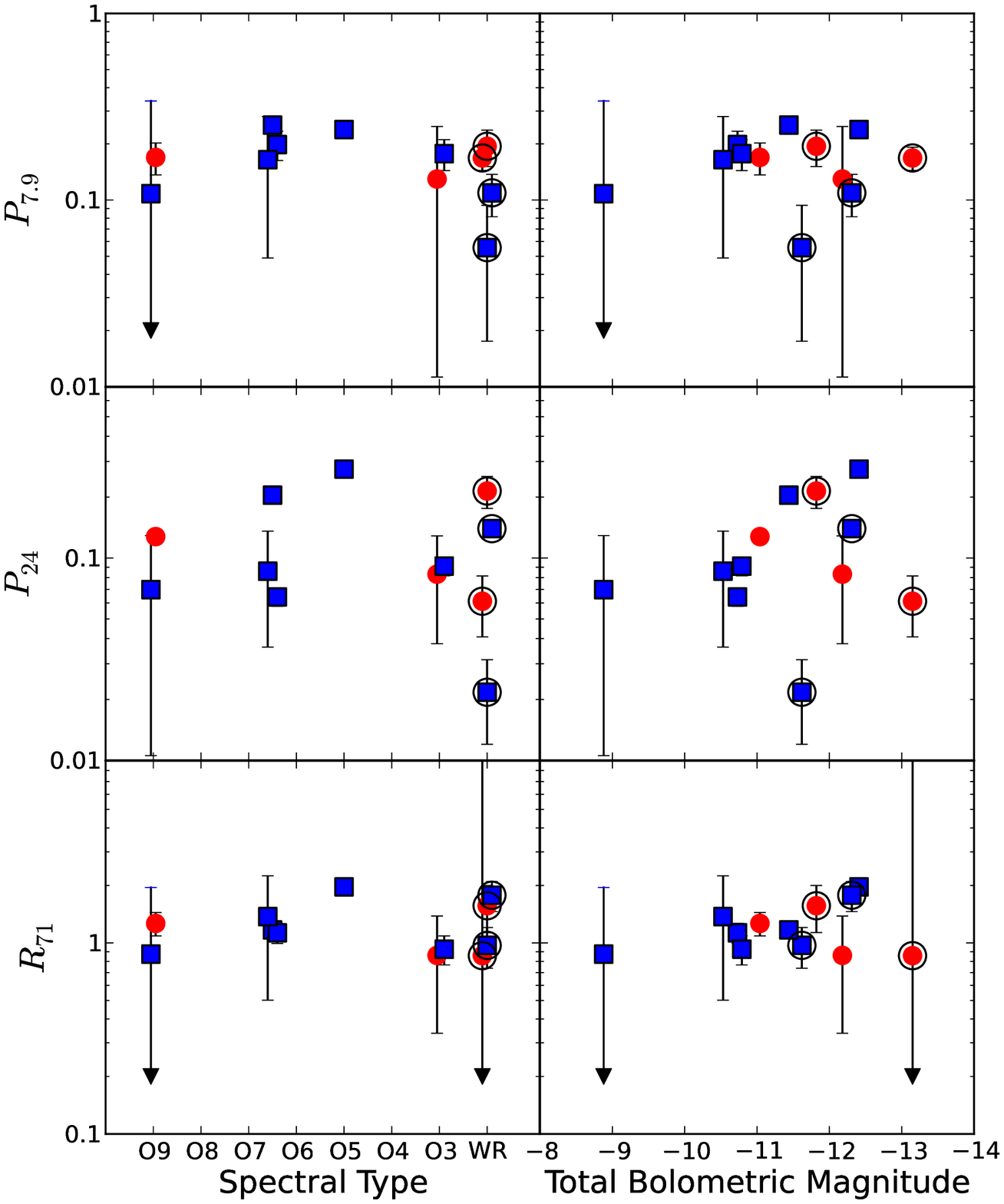}
\caption{The infrared emission ratios are plotted against
  spectral type of the hottest ionizing star in each region ({\it
    left}), and against the total bolometric magnitude of massive
  stars in the region ({\it right}). Superbubbles are marked with
  blue squares, while non-superbubbles are marked with red circles.
  Objects that contain Wolf-Rayet stars are circled. Slight horizontal
  offsets have been added to points in the left panels for clarity.
  \label{sptype}}
\end{figure}

\end{document}